\documentclass[10pt, conference, compsocconf]{IEEEtran}
\IEEEoverridecommandlockouts 
\usepackage{cite}
\usepackage[cmex10]{amsmath}
\usepackage{array}
\usepackage{theorem}
\usepackage[caption=false,font=footnotesize]{subfig}
\usepackage{url}
\usepackage{acronym}
\usepackage[section]{placeins}

\ifCLASSINFOpdf
  \usepackage[pdftex]{graphicx}
\else
  \usepackage[dvips]{graphicx}
\fi

{\theorembodyfont{\rmfamily}
   }
{\theorembodyfont{\rmfamily}
   }
{\theorembodyfont{\rmfamily}
   }
{\theorembodyfont{\rmfamily}
   }
\def\theenumi{\roman{enumi}}
{\theorembodyfont{\rmfamily}
   }
{\theorembodyfont{\rmfamily}
   }

\begin{document}
%
\title{On the Error Detection Capability of Combined LDPC and\\ CRC Codes for Space Telecommand Transmissions
\thanks{This work was supported in part by
the European Space Agency under contract 4000111690/14/NL/FE.}}


\author{\IEEEauthorblockN{Marco Baldi, Nicola Maturo, Giacomo Ricciutelli, Franco Chiaraluce}
\IEEEauthorblockA{DII, Universit\`a Politecnica delle Marche\\
Ancona, Italy\\
Email: \{m.baldi, n.maturo, f.chiaraluce\}@univpm.it, g.ricciutelli@pm.univpm.it}
}


\maketitle

\begin{abstract}
We present a method for estimating the undetected error rate when a cyclic redundancy check (CRC)
is performed on the output of the decoder of short low-density parity-check (LDPC) codes.
This system is of interest for telecommand links, where new LDPC codes have been designed for updating the current standard. 
We show that these new LDPC codes combined with CRC are adequate for complying with the stringent requirements
of this kind of transmissions in terms of error detection.
\end{abstract}


\section{Introduction}
\label{sec:Intro}

In some applications, error detection is at least as important as error correction.
This is the case of space telecommand (TC) links, where an uncorrected error may cause no command execution, when the error is detected, or wrong execution when the error is undetected.

The Space Link Coding and Synchronization Working Group of the Consultative Committee for Space Data Systems (CCSDS) has recently proposed to update the current recommendations \cite{CCSDS2010} through the inclusion of new low-density parity-check (LDPC) codes \cite{CCSDS2015}. In comparison with the code included in the current standard, which is a simple Bose-Chaudhuri-Hocquenghem (BCH) code able to correct $1$ error and to detect $2$ errors\footnote{In this case the code operates according to the so called single error correction (SEC) mode. Alternatively, a triple error detection (TED) mode is admitted where, however, the code is not allowed to correct any error.}, the new codes are characterized by much higher error correction capability. On the other hand, their error detection capability is known only in part, mostly because explicit formulas for the computation of the undetected codeword error rate (UCER) are not available and very long simulations are needed to estimate the UCER numerically.
Moreover, error detection depends on the decoding algorithm: in case of complete decoders the UCER coincides with the codeword error rate (CER), while using incomplete decoders allows to improve the UCER performance at the expense of CER performance.
Indeed, the CER values required in TC links (typically, $\leq 10^{-5}$) are significantly higher than those required for the UCER (typically, $\leq 10^{-9}$) \cite{Chiaraluce2014}, and this may be problematic with complete decoders.

A classical solution to improve the error detection performance consists of adding an outer cyclic redundancy check (CRC) code. The CRC code has no correction capability but it is able to detect a given number of random errors (in addition to some bursts error detection capability). Actually, the TC synchronization and channel coding standard includes, as an option, a $16$-bit CRC code which is able to detected up to $4$ errors \cite{CCSDS2015b}.
Combining the error detection capability of LDPC and CRC codes is gaining an increasing interest \cite{Prodan2013}. 
In order to estimate the overall error detection performance, an approach often used consists of multiplying the UCER at the output of the LDPC decoder, estimated through simulation, by $2^{-P}$, being $P$ the number of redundancy bits of the CRC code.
In fact, the factor $2^{-P}$ represents the average fraction of input sequences that produce the same CRC syndrome, thus resulting indistinguishable one each other and eventually producing an undetected error. 
Therefore, multiplying the LDPC decoder UCER by $2^{-P}$ corresponds to assume that an undetected error pattern at the output of the LDPC decoder may belong, with the same probability, to any syndrome coset.
This assumption is not obvious, and needs verification. So, one of the goals of this paper is to determine the overall UCER performance in a more precise way.

As we will show in the following, the exact estimation of the performance of the CRC code concatenated with the LDPC code requires the knowledge of the weight spectrum of the LDPC code, and this is generally a hard problem. 
For the short LDPC codes proposed for TC links, this problem has been faced and the analysis developed around the subject has already produced valuable results, that will be reminded afterward. 

Indeed, an exhaustive analysis is still not possible. 
However, after validating our approach on a reduced scale, we are able to resort to an accurate approximation.
This way, we can obtain a reliable estimate of the UCER performance for the CRC + LDPC coding scheme in a scenario of practical interest and to compare it with the conventional approach.
This also allows to verify compliance with the error detection requirements of space TC links, which is another important goal of this study.

The organization of the paper is as follows. In Section \ref{sec:Codes} we describe the considered scheme and evaluate its CER/UCER performance in absence of CRC. 
In Section \ref{sec:CRC} we introduce our analysis method and apply it to a toy example as well as to the LDPC codes of interest for TC links. 
Finally, Section \ref{sec:Conclusion} concludes the paper.

\section{Description of the coding scheme}
\label{sec:Codes}

\begin{figure}[t]
\begin{centering}
\includegraphics[width=90mm, keepaspectratio]{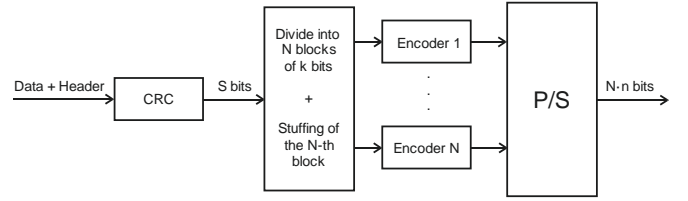}
\caption{\label{fig:Encoder} Block scheme of the considered encoder.}
\par\end{centering}
\end{figure}

The considered encoding scheme is schematically shown in Fig. \ref{fig:Encoder}. 
According to the standard \cite{CCSDS2010}, the $S$ bits at the output of the CRC, with $64 \leq S \leq 8192$, form the payload of a variable-length  transfer frame (TF), which is then divided into $N = \lceil \frac{S}{k} \rceil$ blocks, with $k$ properly chosen.
Stuffing is used to complete the $N$-th block, if necessary. Then, each of these blocks is encoded by using a block code $C(n, k)$, where $n$ is the codeword length. The codewords pattern at the output of the parallel to serial (P/S) converter is sent to another block which adds start and tail sequences, that are introduced for aiding synchronization at the receiver side. These sequences, however, have no role in the present analysis and are therefore omitted.
At the receiver, the scheme is dual, that is, the received pattern is divided into $N$ blocks and each of them is decoded separately from the others. The $S$ bits at the output, after having eliminated the stuffing bits (if present), are then sent to the CRC for integrity checking.

The CRC code is defined by the following generator polynomial \cite{CCSDS2015b}
\begin{equation}
g_{CRC}^{(16)}(x) = x^{16} + x^{12} + x^{5} + 1.
\label{gen_pol_16}
\end{equation}
Regarding $C(n, k)$, the current standard uses a BCH code with $n = 63$ and $k = 56$, which means that the rate is $R = 56/63$. The new LDPC codes, instead, have $R = 1/2$ and $k = 64$ or $k = 256$. In this paper, we mainly focus on the shortest LDPC(128, 64) code, though the analysis could be repeated for the longest one. 

The parity-check matrix of the LDPC(128, 64) code can be obtained starting from the base matrix $\left[\mathbf{A} | \mathbf{B} \right]$, with \cite{CCSDS2015}:
\begin{align}
\mathbf{A} & = \left[ \begin{array}{cccc}
\mathbf{I}_M + \mathbf{\Phi}^7 	& \mathbf{\Phi}^2 														& \mathbf{\Phi}^{14} 												& \mathbf{\Phi}^6 \\
\mathbf{\Phi}^6 													& \mathbf{I}_M + \mathbf{\Phi}^{15} 	& \mathbf{\Phi}^0 														& \mathbf{\Phi}^1 \\
\mathbf{\Phi}^4 													& \mathbf{\Phi}^1 														& \mathbf{I}_M + \mathbf{\Phi}^{15} 	& \mathbf{\Phi}^{14} \\
\mathbf{\Phi}^0 													& \mathbf{\Phi}^1 														& \mathbf{\Phi}^9 														& \mathbf{I}_M + \mathbf{\Phi}^{13} 
\end{array} \right], \nonumber \\
\mathbf{B} & = \left[ \begin{array}{cccc}
\mathbf{0}_M 				& \mathbf{\Phi}^0 	& \mathbf{\Phi}^{13} 	& \mathbf{I}_M \\
\mathbf{I}_M 					& \mathbf{0}_M 			& \mathbf{\Phi}^0 			& \mathbf{\Phi}^7 \\
\mathbf{\Phi}^{11} & \mathbf{I}_M 			& \mathbf{0}_M 					& \mathbf{\Phi}^3 \\
\mathbf{\Phi}^{14} & \mathbf{\Phi}^1 	& \mathbf{I}_M 					& \mathbf{0}_M
\end{array} \right].
\label{H_matrix}
\end{align}
According to (\ref{H_matrix}), the parity-check matrix is formed by $M \times M$ submatrices where $M = k/4 = n/8$.
$\mathbf{I}_M$ and $\mathbf{0}_M$ are the $M \times M$ identity and zero matrices, respectively, and $\mathbf{\Phi}$ is the first right circular shift of $\mathbf{I}_M$. Explicitly, this means that $\mathbf{\Phi}$ has a non-zero entry at row $i$ and column $j$ iff $j = i + 1 \mod M$. Consequently, $\mathbf{\Phi}^2$ is the second right circular shift of $\mathbf{I}_M$, that is, $\mathbf{\Phi}^2$ has a non-zero entry at row $i$ and column $j$ iff $j = i + 2 \mod M$, and so on. Obviously, $\mathbf{\Phi}^0 = \mathbf{I}_M$. The $\oplus$ operator indicates modulo-2 addition.

\begin{figure}[t]
\begin{centering}
\includegraphics[width=85mm, keepaspectratio]{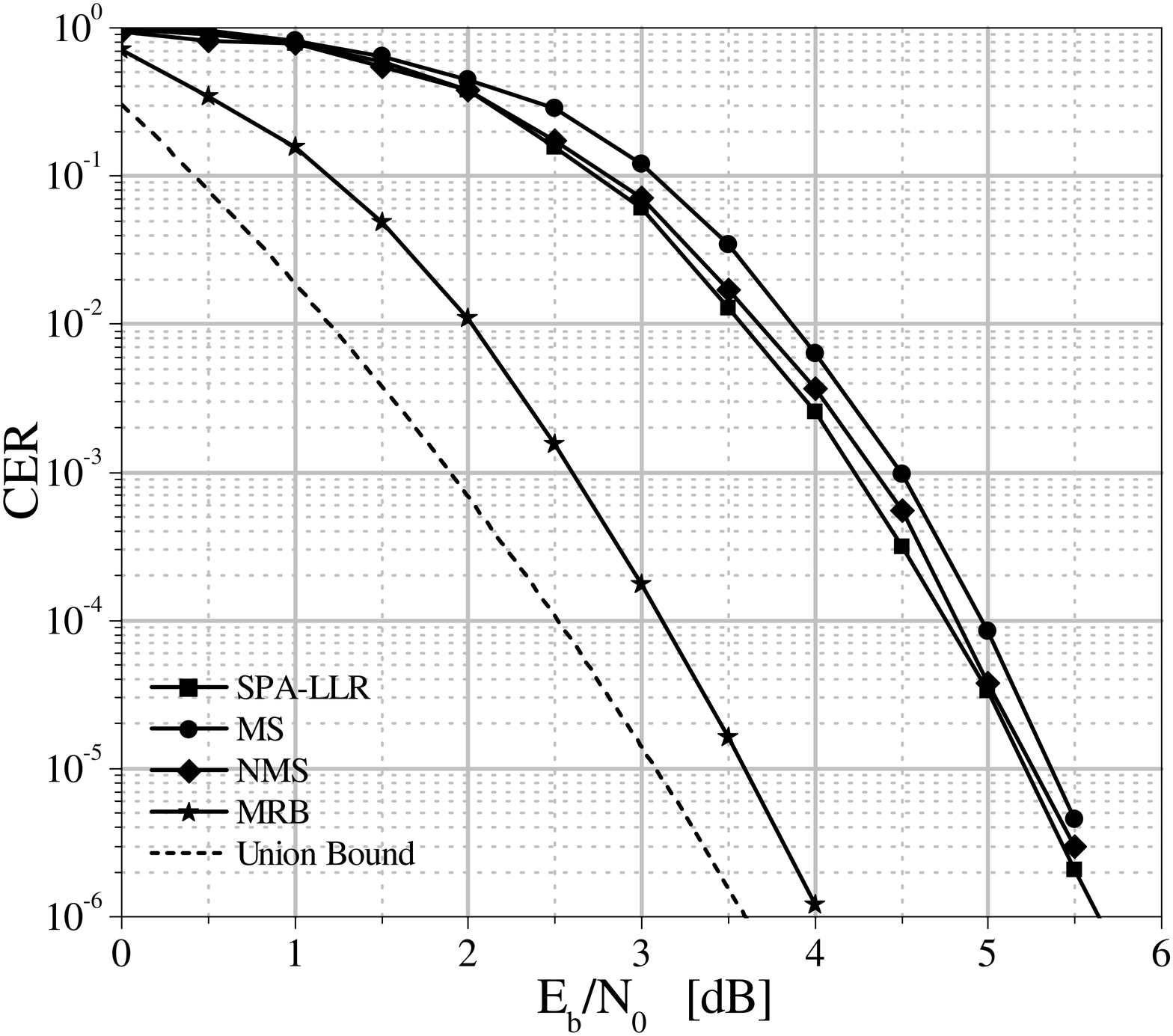}
\caption{\label{fig:CER_128_64} CER performance of the LDPC(128, 64) code with BPSK modulated transmission over the AWGN channel and different decoding algorithms.}
\par\end{centering}
\end{figure}

We have widely investigated the performance of this code over the AWGN channel, by considering binary phase shift keying (BPSK) modulation and using a variety of decoding algorithms. In Fig. \ref{fig:CER_128_64}, for example, we report the performance of three incomplete decoders exploiting iterative algorithms, namely, sum-product algorithm based on log-likelihood ratios (SPA-LLR) \cite{Hagenauer1996}, min-sum (MS) \cite{Fossorier1999} and normalized min-sum (NMS) \cite{Chen2002b}, and one complete decoder, based on the so called most reliable basis (MRB) algorithm \cite{Fossorier1995}. As evident from the figure, the complete decoder provides a significant gain with respect to the incomplete decoders, in terms of CER. Explicitly, the SNR (expressed as the ratio between the energy per bit $E_b$ and the one-side spectral density of the thermal noise $N_0$) required to achieve CER = $10^{-5}$ is $E_b/N_0 \approx 3.6$ dB for the MRB algorithm and $E_b/N_0 \approx 5.2$ dB for the SPA-LLR (and slightly greater for the other algorithms), therefore the gain achieved by MRB decoding is in the order of $1.6$ dB. 
The figure also includes the so called union bound curve. This curve provides an upper bound on the error rate of the considered code under maximum likelihood (ML) decoding. The expression of the union bound for the CER (a similar expression can be derived for the bit error rate (BER)) is as follows \cite{Proakis1995}
\begin{equation}
{\rm CER_{UB}} = \sum_{w=d_{\min}}^{n} \frac{1}{2}A_w {\rm erfc} \sqrt{w R \frac{E_b}{N_0}}
\label{eq:UB_CER}
\end{equation}
where $A_w$ is the weight-$w$ multiplicity, that is, the number of codewords with weight $w$, and $d_{\min}$ is the minimum (Hamming) distance of the code $C(n, k)$. The first term of the sum, corresponding to $w = d_{\min}$, is also known as the ``error floor''. For sufficiently high values of $E_b/N_0$ it provides an excellent approximation of the performance of ML decoding.

The evaluation of the union bound requires the knowledge of the weight spectrum of the code. It is known that for LDPC codes this may be a non-trivial task. For the LDPC(128, 64) code, however, much work has been done to circumvent this issue. In particular, the first and most significant terms of the weight distribution for the LDPC(128, 64) code are specified in polynomial form as
\begin{align}
\left. A(x) \right|_{64 \times 128} &= 16x^{14} + 528x^{16} + 5632x^{18} \notag \\ 
                                                        &+ 35968x^{20} + 123888x^{22} + 364944x^{24} + \dots
\label{eq:EWFs}
\end{align}
where the presence of the term $A_w\, x^w$ means that there are $A_w$ codewords with Hamming weight $w$. The multiplicities $A_{14}$, $A_{16}$ and $A_{18}$ are exact \cite{JPL2015}; this part of the weight spectrum has been obtained through computer searches using a carefully tuned ``error impulse'' technique \cite{Declercq2008}. The other multiplicities are lower bounds on the actual values and have been obtained by using the approach proposed in \cite{Hu2004a}. It should be noted that any multiplicity can be expressed as a multiple of $16$. This is due to the quasi-cyclic structure of the code, according to which any codeword can be seen as the concatenation of $\frac{n}{M} = 8$ blocks, each one consisting of $M$ bits. Any cyclic shift by $1, 2, ..., M - 1$ positions within all blocks of a codeword (block-wise cyclic shift) produces another codeword. Explicitly, also in view of the application of the method we will propose next, this means that, for the LDPC(128, 64) code with $M = 16$, codewords are found in groups of $16$.

From Fig. \ref{fig:CER_128_64} we see that, also in comparison with the asymptotic ML behavior, the CER performance of the  MRB algorithm is very good: the gap with respect to the union bound curve is about $0.5$ dB at CER = $10^{-5}$. On the opposite, the iterative algorithms are significantly suboptimal for the considered code.
On the other hand, if we pass to consider the UCER performance, the situation is reversed.

Let us consider first the SPA-LLR decoder. During simulations, undetected errors occur as a subset of the whole ensemble of errors. In the expected event that the UCER assumes very low values, at the SNR of interest, very long simulations are required to find a statistically significant number of undetected errors. More precisely, noting by $Q_u$ the number of undetected errors found at the output of a simulation that has produced $Q$ errors in total, the undetected error rate can be estimated as ${\rm UCER} = \frac{Q_u}{Q}{\rm CER}$. It is evident that, in order to have a sufficiently high statistical confidence, the value of $Q_u$ must be sufficiently large and, for such a goal, very long simulations are required.

\begin{figure}[t]
\begin{centering}
\includegraphics[width=85mm, keepaspectratio]{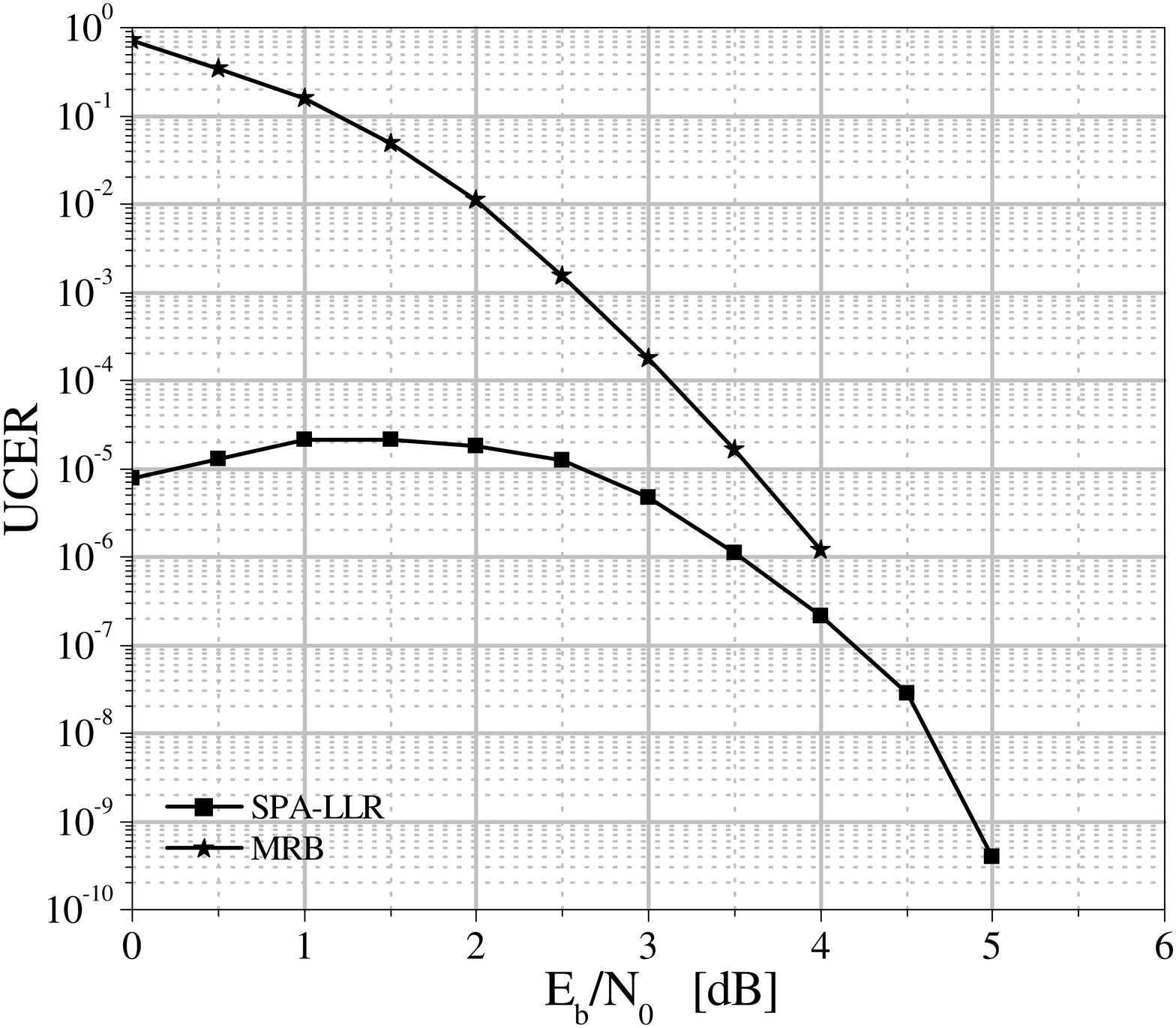}
\caption{\label{fig:UCER_128_64} UCER performance of the LDPC(128, 64) code with BPSK modulated transmission over the AWGN channel and SPA-LLR and MRB decoding.}
\par\end{centering}
\end{figure}

The UCER curve of the LDPC(128, 64) code, under SPA-LLR decoding is shown in Fig. \ref{fig:UCER_128_64}. This curve has been obtained by imposing to find $500,000$ erred codewords, as the stopping rule, that is $5,000$ times greater than the standard criterion for example adopted to obtain the CER curves of Fig. \ref{fig:CER_128_64}. In both cases, the maximum number of iterations used was $I_{max} = 100$. We see that UCER = $10^{-9}$ is reached when $E_b/N_0 \approx 4.8$ dB, that is a value smaller than the working point ($E_b/N_0 \approx 5.2$ dB) required to satisfy the requirement on the CER.

The figure also shows the UCER curve when using the MRB algorithm. Actually, this curve coincides with that reported in Fig. \ref{fig:CER_128_64} since, as mentioned, UCER = CER for this scheme. The curve does not permit to estimate the SNR required to achieve UCER = $10^{-9}$. However, by artificially prolonging the curve, under the reasonable assumption that no error floor appears, we can foresee that the SNR required to satisfy the requirement on the UCER cannot be smaller than $E_b/N_0 \approx 5$ dB.
This value is significantly larger than the working point ($E_b/N_0 \approx 3.6$ dB) required to satisfy the requirement on the CER with the same decoding algorithm. This means that the potential advantage offered by the MRB algorithm may be frustrated and, wishing to satisfy both constraints (that is, on the CER and the UCER) the system should operate at an SNR comparable with that needed by the SPA-LLR decoder.

In such a scenario, the CRC can play a fundamental role, contributing to improve the UCER performance of the LDPC decoder. For the reasons explained above, the CRC is ``mandatory'' when adopting the MRB algorithm. However, it is certainly useful also when employing iterative algorithms.
So, in the following section we introduce a method for estimating the performance of the outer CRC code which is alternative to the rough approach reminded in the Introduction, and we provide relevant numerical results.

\section{Analysis of the CRC code performance}
\label{sec:CRC}

For the sake of simplicity, let us start by considering $S = 64$ bits, so that the encoded TF consists of only one codeword ($N = 1$). This assumption will be removed in Section \ref{subsec:Extension}. However, it is important to note that it corresponds to a situation of practical interest, occurring in case of short emergency commands.

Let us denote by $[\mathbf{s}|\mathbf{p}]$ the information ($\mathbf{s}$) plus CRC redundancy ($\mathbf{p}$) vector, of size $k$, representing the correct sequence and by $[\mathbf{s^*}|\mathbf{p^*}]$ an erred one resulting from an undetected error at the output of the LDPC decoder.
In the rare cases in which $\mathbf{p^*}$ actually corresponds to the CRC redundancy computed on $\mathbf{s^*}$, the error is undetected by the CRC code as well.

Let us denote by $c(x)$ the polynomial representing the erred codeword at the output of the LDPC decoder. The polynomial representing $[\mathbf{s^*}|\mathbf{p^*}]$, i.e., the first part of $c(x)$, can be written as $m(x) = \sum_{i=0}^{k - 1} a_i x^i$, $a_i \in[0, 1]$. If the remainder of the division of $m(x)$ by $g_{CRC}(x)$ is zero, then the CRC syndrome check is successful and the error remains undetected; otherwise, the error is detected by the CRC. We call such a procedure ``divisibility test''. In principle, $m(x)$ can assume all possible configurations, that is, any possible combination of powers of $x$.
Actually, We will show next that:
\begin{itemize}
\item the codewords at the output of the LDPC decoder when an undetected error occurs have weights concentrated in the neighborhood of a generally low value which depends on the decoding algorithm,
\item the divisibility of $m(x)$ by $g_{CRC}(x)$ depends on the weight of $c(x)$.
\end{itemize}
The second statement does not agree with the assumption of uniformity, which is at the basis of the multiplying factor $2^{-P}$ mentioned in the Introduction, and influences the value of the UCER at the CRC output. The remark on the weight of the codewords in case of undetected error is particularly important and, to the best of our knowledge, this is the first time such a phenomenon is explicitly observed.


Let us denote by ${\rm{UCER}}_{{\rm{LDPC}}_j}$ the value of the UCER at the output of the LDPC decoder due to weight-$j$ codewords (which means that the erred codewords have weight $j$) and by ${\rm{UCER}}_{{\rm{CRC}}_j}$ the undetected error rate of these codewords by the CRC. The latter is, therefore, a conditional probability. The overall UCER, resulting from the concatenation of the LDPC decoder and the CRC is given by the following expression
\begin{equation}
{\rm UCER} = \sum_{j=d_{\min}}^{n} {\rm{UCER}}_{{\rm{LDPC}}_j} \times {\rm{UCER}}_{{\rm{CRC}}_j}.
\label{eq:UCER_TOT}
\end{equation}
The values of ${\rm{UCER}}_{{\rm{LDPC}}_j}$ can be estimated through long simulations, while those of ${\rm{UCER}}_{{\rm{CRC}}_j}$ can be determined via the polynomial division described above. More precisely, noting by $L_j$ the number of weight-$j$ codewords for which $m(x)$ is divisible by $g_{CRC}(x)$, we can write
\begin{equation}
{\rm{UCER}}_{{\rm{CRC}}_j} = \frac{L_j}{A_j}
\label{eq:UCER_CRC}
\end{equation}
where, according to (\ref{eq:EWFs}), $A_j$ is the weight-$j$ multiplicity.

\subsection{Application to a simple code.}
\label{subsec:Toy_example}
The method described above can be preliminarily applied to an LDPC(32, 16) code. This very short code has been designed by following the same approach, based on protographs, used for the codes to be included in the new standard \cite{CCSDS2015} and described in Section \ref{sec:Codes}. It is obviously not significant in the framework of TC links. However, it permits us to perform an exhaustive search of its codewords (whose number is $2^{16} = 65,536$), this way allowing a complete and rigorous evaluation of the terms ${\rm{UCER}}_{{\rm{CRC}}_j}$ in (\ref{eq:UCER_CRC}).

The generator polynomial of the CRC code for this example is assumed to be
\begin{equation}
g_{CRC}^{(8)}(x) = x^{8} + x^{7} + x^{6} + x^{4} + x^{2} + 1.
\label{gen_pol_8}
\end{equation}

The complete, exact weight distribution of the code is summarized in the following polynomial
\begin{align}
\left. A(x) \right|_{16 \times 32} &= 4x^{4} + 48x^{6} + 460x^{8} \notag \\ 
                                                        &+ 1776x^{10} + 6684x^{12} + 14048x^{14} + 19494x^{16} \notag \\
																												&+ 14048x^{18} + 6684x^{20} + 1776x^{22} + 460x^{24} \notag \\
																												&+ 48x^{26} + 4x^{28} + x^{32}.
\label{eq:EWFs_toy}
\end{align}

\begin{figure}[t]
\begin{centering}
\includegraphics[width=85mm, keepaspectratio]{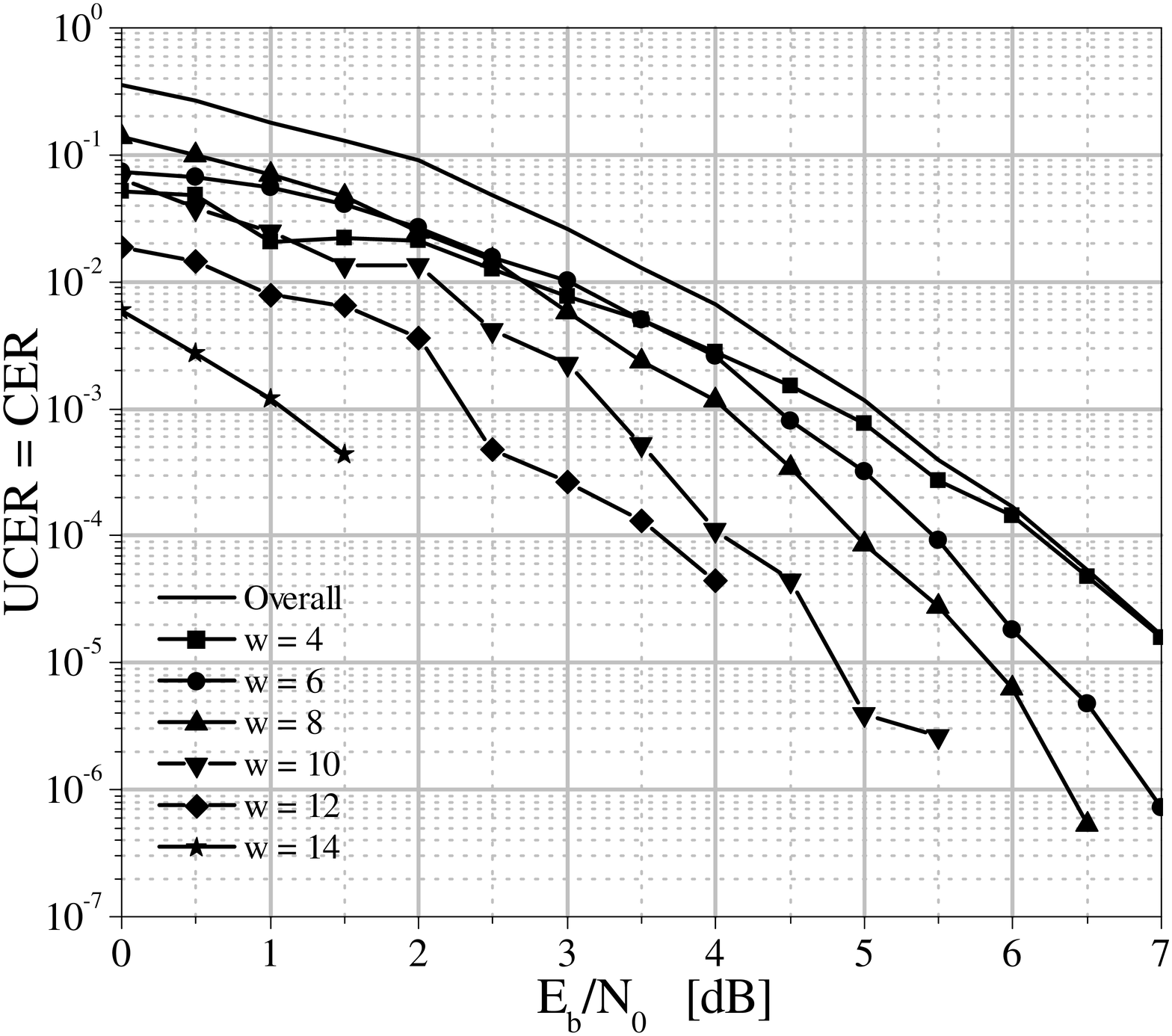}
\caption{\label{fig:UCER_32_16_MRB} UCER performance of the LDPC(32, 16) code, with detail of individual contributions, by using the MRB algorithm.}
\par\end{centering}
\end{figure}

Let us focus on the MRB decoding algorithm, that we use with order $4$ (see \cite{Baldi2015} for details). A similar analysis can be developed for the SPA-LLR, but it is here omitted for saving space. In Fig. \ref{fig:UCER_32_16_MRB} we report the UCER curve (coincident, as usual, with the CER curve) decomposed into the contributions due to codewords of different weights. These curves are the result of a Montecarlo simulation; hence, they must be considered as an estimate. In particular, each curve is interrupted at the value of $E_b/N_0$ above which simulation has not found erred codewords with the specified weight. Weights larger than $14$ do not appear, for the same reason. However, their incidence, particularly for not too small values of $E_b/N_0$, is expected to be quite negligible. Indeed, we see that for higher and higher SNR, the major contribution to the UCER comes from the smallest weight codewords (those with weight $4$, in particular).

As the weight spectrum of this code is completely known, it is possible to compute exactly the terms $L_i$. Those non null are reported in Table \ref{tab:Li_32_16}. For $i = 4, 6, 8, 26, 28, 32$ we have in fact $L_i = 0$ which means that undetected erred codewords with such weights, at the output of the LDPC decoder, are certainly identified by the CRC and do not contribute to the overall UCER at the receiver output. This is very important since, as seen in Fig. \ref{fig:UCER_32_16_MRB}, the low-weight codewords are responsible for the major contributions to the UCER at the output of the LDPC decoder, that will be, therefore, significantly smoothed in the presence of the CRC.
\begin{table}[t]
\renewcommand{\arraystretch}{1.1}
\caption{Non null values of $L_i$ for the LDPC(32, 16) code.}
\label{tab:Li_32_16}
\centering
\scriptsize
\begin{tabular}{|c|c|c|c|c|c|c|c|c|}
\hline
$i$ & $10$ & $12$ & $14$ & $16$ & $18$ & $20$ & $22$ & $24$\\
\hline
$L_i$ & $9$ & $26$ & $52$ & $72$ & $61$ & $28$ & $6$ & $1$\\
\hline
\end{tabular}
\end{table}

\begin{figure}[t]
\begin{centering}
\includegraphics[width=85mm, keepaspectratio]{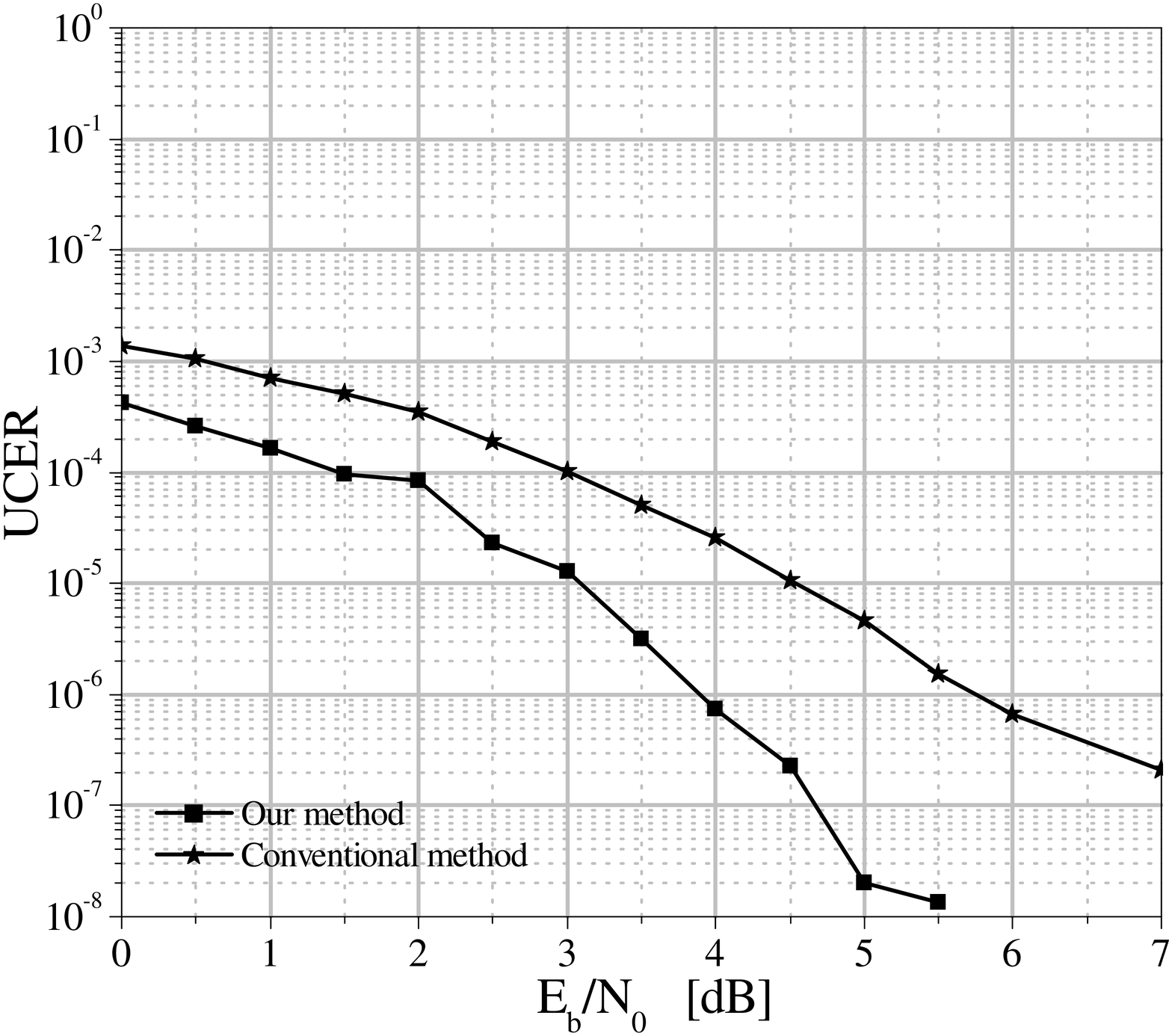}
\caption{\label{fig:UCER_overall_short} Estimated overall UCER for the LDPC(32, 16) code with MRB decoding and comparison with the result obtained by the conventional method.}
\par\end{centering}
\end{figure}

Actually, from the values of $L_i$, the ${\rm{UCER}}_{{\rm{CRC}}_j}$'s can be determined through (\ref{eq:UCER_CRC}). Finally, multiplying by the corresponding ${\rm{UCER}}_{{\rm{LDPC}}_j}$ and summing, according to (\ref{eq:UCER_TOT}), the UCER curve after application of the LDPC (MRB) decoder and the CRC is that shown in Fig. \ref{fig:UCER_overall_short}. The curve stops at $E_b/N_0 = 5.5$ dB, as this is the last simulated point for the codewords with weight $i = 10$, which are the first to provide a non null contribution. The figure also shows a comparison with the curve obtainable by using the conventional method, as described in Section \ref{sec:Intro}, which consists of multiplying ${\rm{UCER}}_{\rm{LDPC}}$ by $2^{-P} = 2^{-8}$. We see that the difference between the two curves is significant: for this particular code, the conventional method overestimates the UCER by about two orders of magnitude at high SNRs, which are the most interesting in practice.

\subsection{Application to the LDPC(128, 64) code.}
\label{subsec:LDPC_example}
In principle, application of the method to the LDPC(128, 64) code proceeds exactly as described for the shorter code in Section \ref{subsec:Toy_example}. In this case, however, the number of codewords is $2^{64} \approx 1.8 \cdot 10^{19}$ and this prevents us from finding all codewords in an acceptable time, as required by the divisibility test. More precisely, we were able to determine all codewords with weight $i = 14, 16$ and $18$ (the same for which the multiplicities in (\ref{eq:EWFs}) are exact), while for the others we have been able to estimate a subset that, however, is significant enough for our evaluation.

\begin{figure}[t]
\begin{centering}
\includegraphics[width=60mm, keepaspectratio]{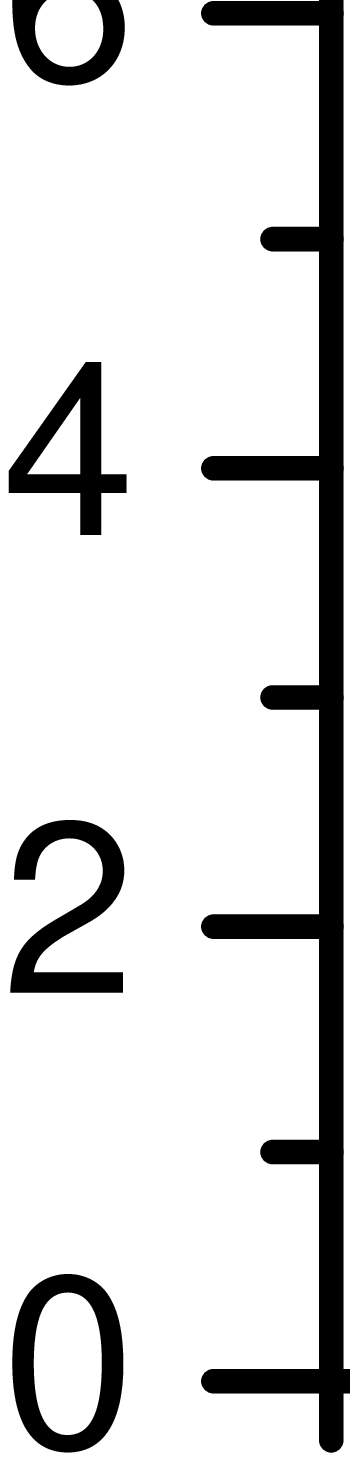}
\includegraphics[width=60mm, keepaspectratio]{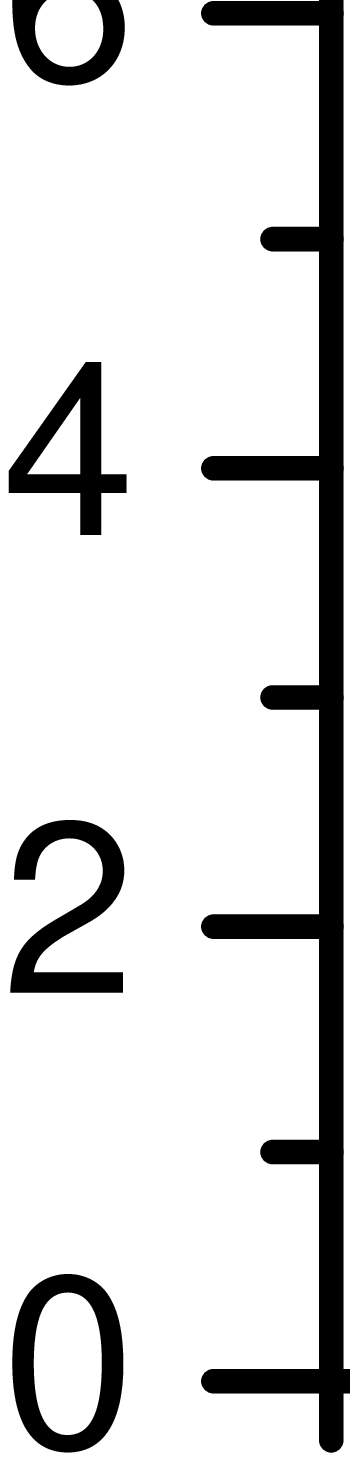}
\includegraphics[width=60mm, keepaspectratio]{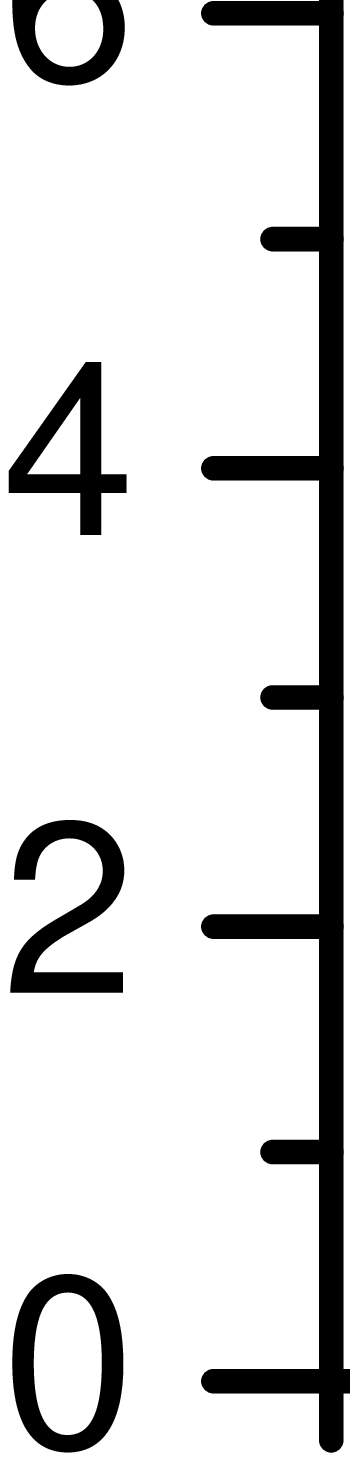}
\caption{\label{fig:Histograms_MRB} Multiplicities of erred codewords at the output of the MRB decoder.}
\par\end{centering}
\end{figure}

Let us suppose to perform decoding by MRB of order $4$. Figure \ref{fig:Histograms_MRB} shows the number of erred codewords we have found through simulation, as a function of the codewords weight, for different values of $E_b/N_0$, at the output of the MRB decoder. For each SNR point, we have run simulations until finding $100$ erred codewords (and, in fact, the columns of the histogram sum to $100$). This guarantees a satisfactory confidence level for the error rates here considered. Differently from Fig. \ref{fig:UCER_32_16_MRB}, we have preferred to decompose the contributions due to different weight codewords by looking at the multiplicities instead of the error rate, as this helps to discuss results. Attention has been focused on the interval $E_b/N_0 \in [3, 4]$ dB since, according to Fig. \ref{fig:CER_128_64}, this is the region of interest in view of satisfying the constraint on the CER.
From the figures we see that, similarly to the shorter code in Section \ref{subsec:Toy_example}, by increasing the value of $E_b/N_0$, the majority of the erred codewords tend to have lower and lower weights. Actually, this mechanism becomes even more evident by considering $E_b/N_0 > 4$ dB; the corresponding histograms are not reported here for the sake of brevity. We can conclude that if the SNR value is sufficiently high, the ${\rm{UCER}}_{\rm{LDPC}}$ curve is dominated by these low weight codewords, and the estimation error is acceptable, in the neighborhood of $E_b/N_0 = 3.5$ dB, by considering weights up to $28$.

As mentioned above, in this case we are not able to find the complete $L_i$-distribution, as we do not know the complete set of codewords. However, by using the subset of codewords we have found, we have been able to establish that $L_{14} = L_{16} = L_{18} = 0$. This result is exact, as for these weights the divisibility test was realized exhaustively. Accordingly, the sum in (\ref{eq:UCER_TOT}) can start from $j = 20$. On the other hand, for the reasons explained above, we can truncate the sum at $j = 28$, at the cost of an acceptable error.

\begin{figure}[t]
\begin{centering}
\includegraphics[width=85mm, keepaspectratio]{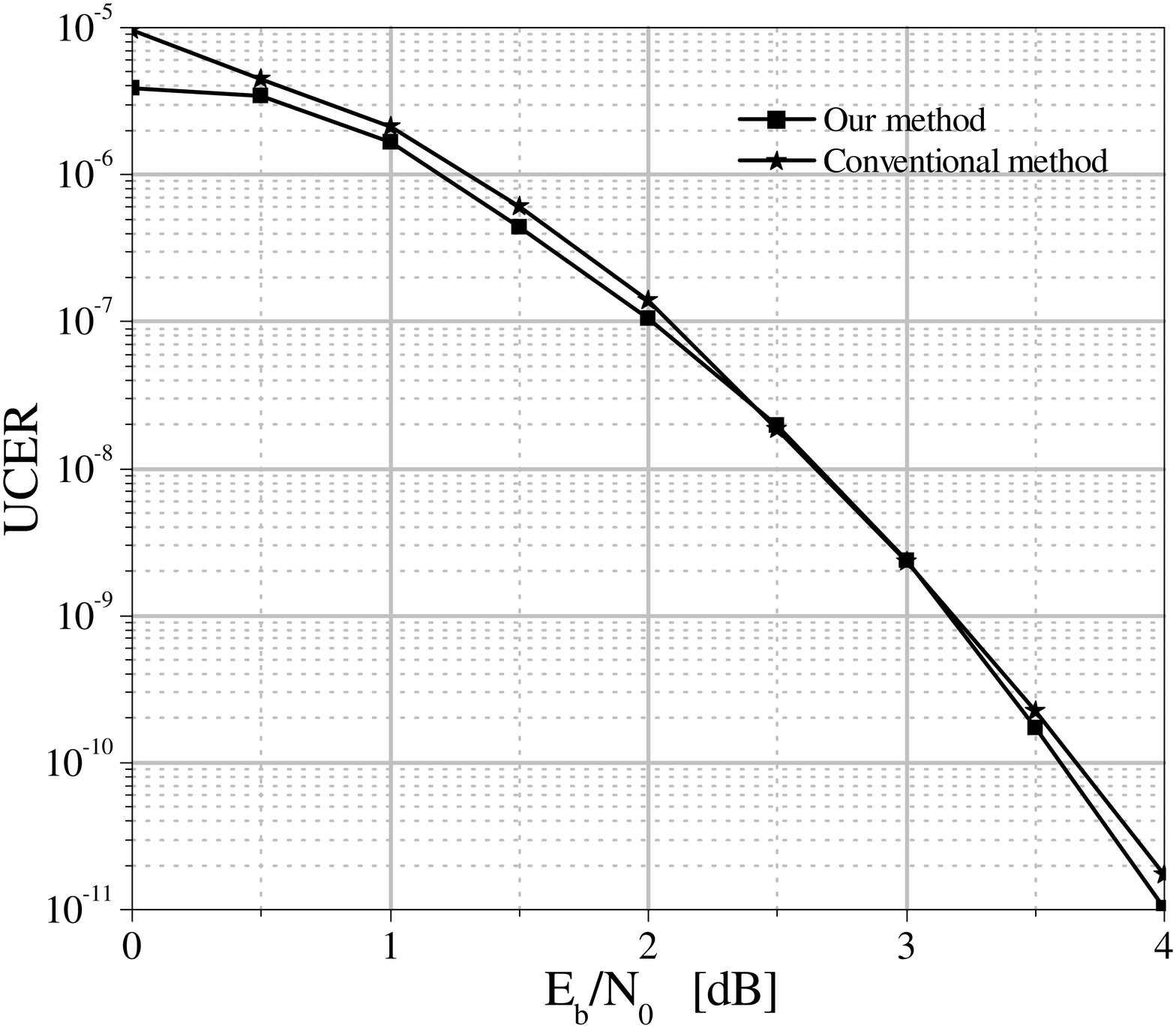}
\caption{\label{fig:UCER_overall_MRB} Estimated overall UCER for the LDPC(128, 64) code, with MRB decoding, and comparison with the result obtained by the conventional method.}
\par\end{centering}
\end{figure}

Following this strategy, we have estimated the values of $L_j$, on the basis of the available codewords, for $j = 20, 22, 24, 26$ and $28$, replaced them in (\ref{eq:UCER_CRC}) (where the $A_j$ values are approximate as well), and finally combined the values of ${\rm{UCER}}_{{\rm{CRC}}_j}$ so obtained with those of ${\rm{UCER}}_{{\rm{LDPC}}_j}$. This way, we have been able to derive a meaningful estimate of the overall UCER at the receiver output.
The result of such processing is shown in Fig. \ref{fig:UCER_overall_MRB} where, similarly to Fig. \ref{fig:UCER_overall_short}, a comparison has been done with the curve obtained by applying the conventional method, that here consists of multiplying ${\rm{UCER}}_{\rm{LDPC}}$ by $2^{-P} = 2^{-16}$. We see that, contrary to the LDPC(32, 16) code, in this case our method provides a curve very close to that achievable by applying the conventional method. Therefore, the latter can be applied with good confidence.

From a practical point of view, the most important conclusion we can draw from Fig. \ref{fig:UCER_overall_MRB} is that concatenation of the LDPC MRB decoder and the CRC allows to satisfy the constraint on the UCER at the working point fixed by the CER; at $E_b/N_0 = 3.6$  dB, in fact, we have ${\rm{UCER} \approx 10^{-10}}$. Actually, this conclusion could be drawn directly by applying the conventional method but the fact to have proven it by exploiting a more rigorous approach makes the analysis more convincing.

\begin{figure}[t]
\begin{centering}
\includegraphics[width=60mm, keepaspectratio]{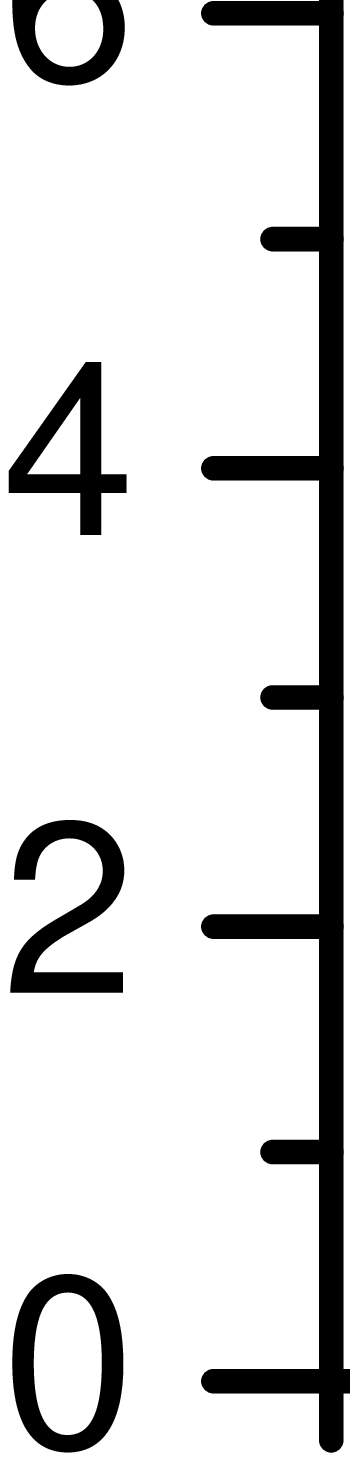}
\includegraphics[width=60mm, keepaspectratio]{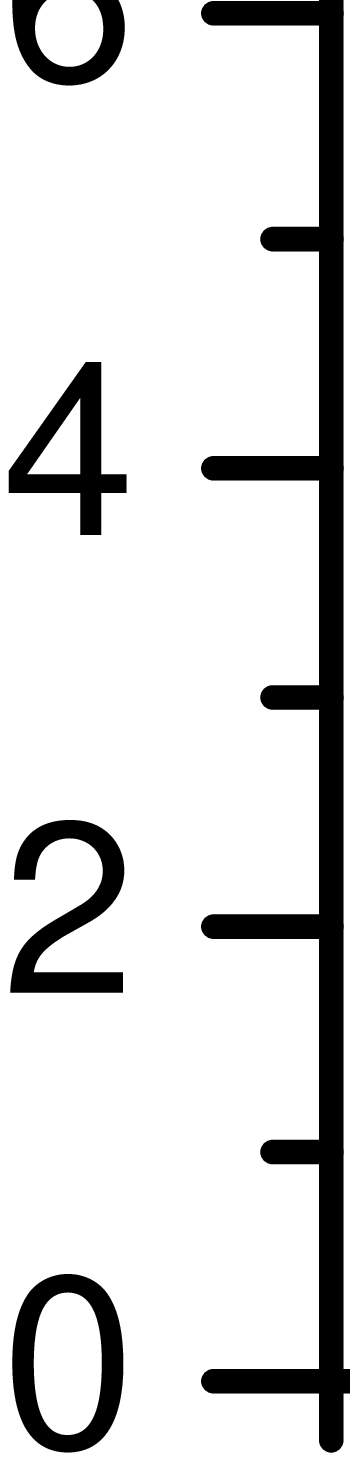}
\includegraphics[width=60mm, keepaspectratio]{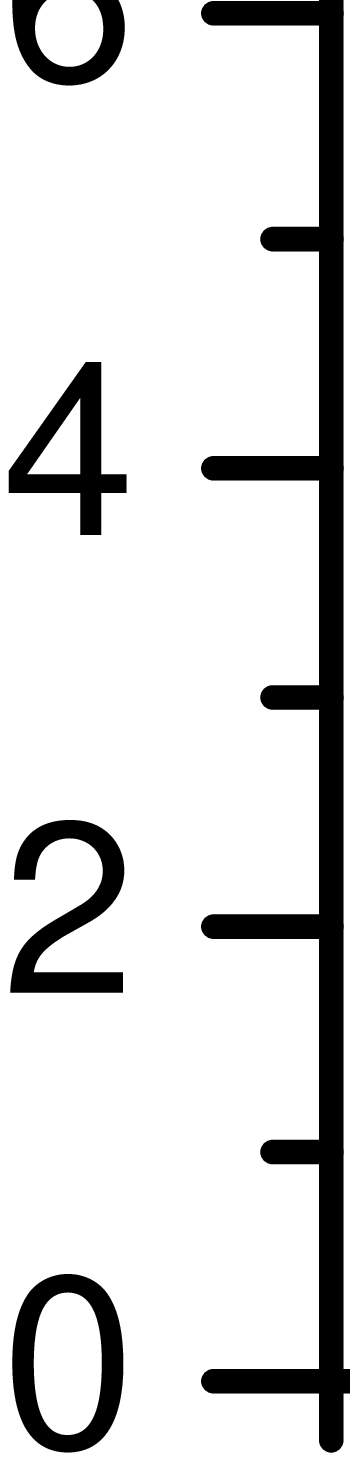}
\caption{\label{fig:Histograms_SPA} Multiplicities of erred codewords at the output of the SPA-LLR decoder.}
\par\end{centering}
\end{figure}

As seen in Fig. \ref{fig:UCER_128_64}, when the SPA-LLR is used for decoding the LDPC(128, 64) code the CRC is unnecessary, as the UCER requirement is satisfied at $E_b/N_0 = 5.2$ dB, that is the working point fixed by the constraint on the CER. However, the CRC can further lower the value of the UCER, this way increasing the margin with respect to the fixed requirement. So, the analysis developed above in case of applying the MRB algorithm has been repeated for the SPA-LLR. Figure \ref{fig:Histograms_SPA} shows the number of erred codewords we have found through simulation, as a function of the codewords weight, for values of $E_b/N_0 \in [4, 5]$ dB, at the output of the SPA-LLR decoder. Even more than in Fig. \ref{fig:Histograms_MRB} we see that the weights of the erred codewords tend to be concentrated around the smallest values. Despite the fact that the multiplicities have been found by simulating as large numbers of codewords as to have $500,000$ erred codewords in total, the number of undetected errors, particularly at $E_b/N_0 = 5$ dB, is very small.
However, even excluding the last point, the mentioned trend towards lower and lower weights is confirmed by the other values of $E_b/N_0$ (where the statistical confidence is higher).

\begin{figure}[t]
\begin{centering}
\includegraphics[width=85mm, keepaspectratio]{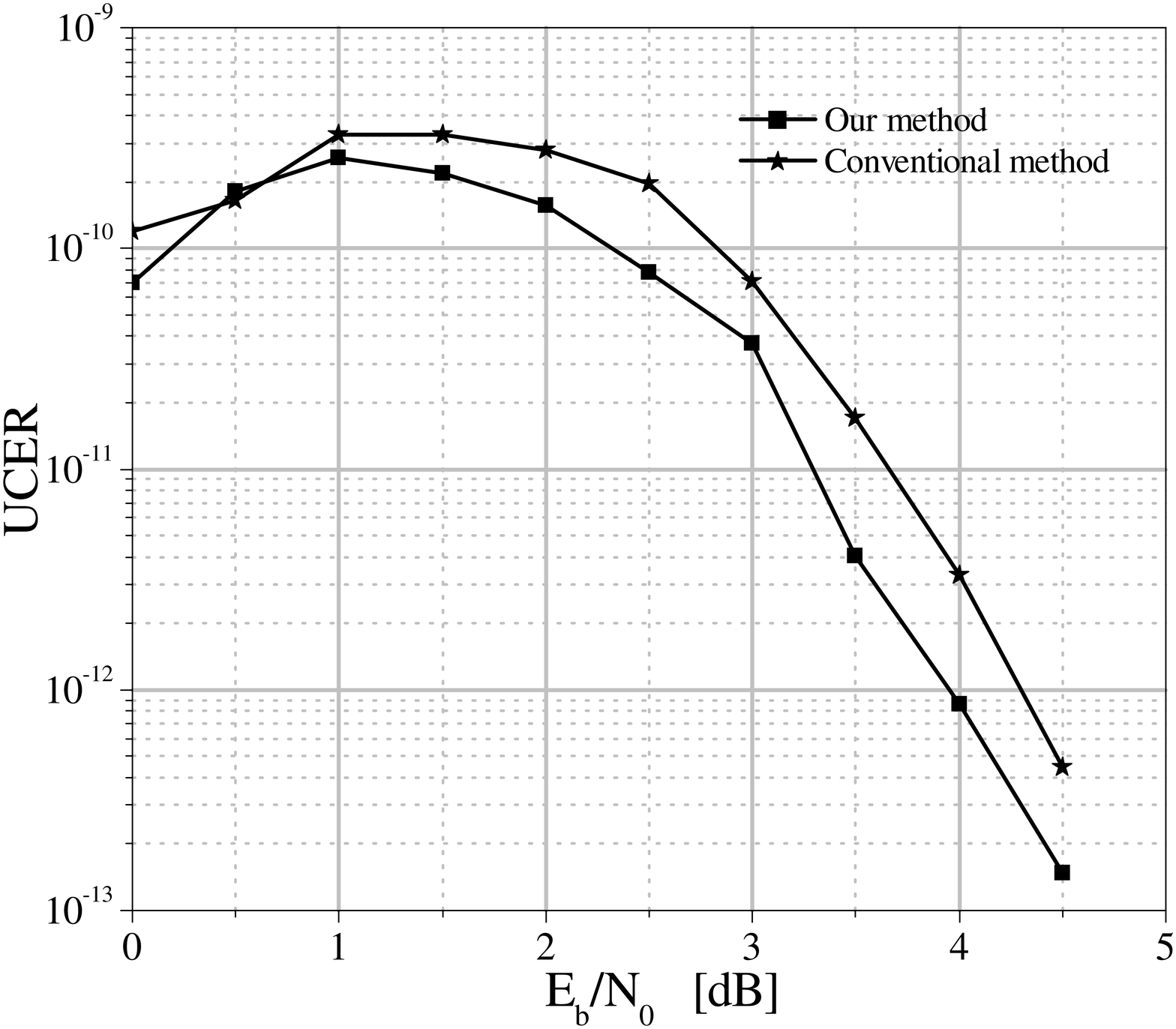}
\caption{\label{fig:UCER_overall_SPA} Estimated overall UCER for the LDPC(128, 64) code, with SPA-LLR decoding, and comparison with the result obtained by the conventional method.}
\par\end{centering}
\end{figure}

By combining the values of ${\rm{UCER}}_{{\rm{LDPC}}_j}$, obtainable from Fig. \ref{fig:Histograms_SPA} (and similar for smaller $E_b/N_0$), with those of ${\rm{UCER}}_{{\rm{CRC}}_j}$, according to (\ref{eq:UCER_CRC}), we obtain the curve in Fig. \ref{fig:UCER_overall_SPA} which contains also the comparison with the conventional method. Contrary to Fig. \ref{fig:UCER_overall_MRB}, the result obtained through our method in this case is appreciably different from that obtained through the conventional method. Both curves do not reach $E_b/N_0 = 5.2$ dB because of the lack of statistical confidence of the simulation results. However, there is no doubt that the concatenation of the LDPC SPA-LLR decoder and the CRC allows to satisfy the constraint on the UCER at the working point fixed by the CER. 

\subsection{Extension to the case of $N > 1$.}
\label{subsec:Extension}
The analysis developed in the previous sections assumed $N = 1$. The conclusions drawn, however, remain basically unchanged even when the encoded TF consists of more than one codeword.

Let us consider the LDPC(128, 64) code. 
In the most general case, the information sequence input to LDPC encoding before transmission over the channel can be written as the vector $[\mathbf{s}_1|\mathbf{s}_2|...|\mathbf{s}_N|\mathbf{p}]$, where each sub-vector $\mathbf{s}_i$, with $i = 1, 2,. . ., N - 1$, has size $k$, while $\mathbf{s}_N$ has size $k - 16$ and $\mathbf{p}$ is the CRC syndrome computed over all previous bits.
After transmission and LDPC decoding, one or more of these sub-vectors may be erred, depending on the decodings that failed. 
However, the probability that two or more decodings result in undetected errors is obviously much smaller than the probability that a single undetected error event occurs.
So, it is realistic to consider the contribution due to a single failure as dominant. 
The probability of this event has been computed in the previous sections for $N = 1$. 
So, the previous analysis, under the assumption of a single failure, strictly holds when the error occurs in $\mathbf{s}_N$.

We remind that the divisibility test, which is at the basis of our method, consists of dividing the polynomial representing the information vector resulting from LDPC decoding by the generator polynomial of the CRC. 
Let us suppose that LDPC decoding of the $N$-th codeword incurred in an undetected error, thus producing $[\mathbf{s}_1|\mathbf{s}_2|...|\mathbf{t}_N]$ as the information vector at the output of LDPC decoding,
with $\mathbf{t}_N \neq [\mathbf{s}_N|\mathbf{p}]$ being the information part of the $N$-th codeword. So, we can write
\begin{equation}
[\mathbf{s}_1|\mathbf{s}_2|...|\mathbf{t}_N] = [\mathbf{s}_1|\mathbf{s}_2|...|\mathbf{s}_N|\mathbf{p}] + [\mathbf{0}|\mathbf{0}|...|\mathbf{e}_N]
\label{eq:gen1}
\end{equation}
where $\mathbf{0}$ is the null vector of size $k$ and $\mathbf{e}_N = \mathbf{t}_N + [\mathbf{s}_N|\mathbf{p}]$ (plus is justified by the fact we are considering binary transmissions). Noting by $p(x)$ the polynomial representing the first vector at the right hand side (r.h.s.) of (\ref{eq:gen1}) and by $m(x)$ the polynomial representing the second vector, $p(x)$ certainly divides $g_{CRC}^{(16)}(x)$. So, in order to check if the CRC is able (or not) to detect the error, it is sufficient to check if $m(x)$ is not divisible (or divisible).

If the error is due to the $j$-th decoder, with $j < N$, in place of (\ref{eq:gen1}) we have
\begin{equation}
[\mathbf{s}_1|...|\mathbf{t}_j|...|\mathbf{s}_N|\mathbf{p}] = [\mathbf{s}_1|...|\mathbf{s}_j|...|\mathbf{s}_N|\mathbf{p}] + [\mathbf{0}|...|\mathbf{e}_j|...|\mathbf{0}]
\label{eq:gen2}
\end{equation}
being $\mathbf{e}_j = \mathbf{t}_j + \mathbf{s}_j$.
The polynomial representing the second vector at the r.h.s. of (\ref{eq:gen2}), noted by $m'(x)$, can be obtained from an $m(x)$ as $m'(x) = x^{64(N - j)} \cdot m(x)$.
Since division of $x^{64(N - j)}$ by $g_{CRC}^{(16)}(x)$ does not produce a remainder equal to zero, it is clear that $m'(x)$ is divisible by $g_{CRC}^{(16)}(x)$ iff $m(x)$ is divisible by $g_{CRC}^{(16)}(x)$.
Therefore, the UCER performance is independent of the position of the erred codeword, and the results obtained in the previous sections are also valid in the most general case.

\section{Conclusions}
\label{sec:Conclusion}
When a CRC code is used as the outer code in a concatenated scheme, evaluation of its performance must take into account the statistical and structural features of the codewords at the output of the inner decoder. By considering the case of short inner LDPC codes, in this paper we have presented a conceptually simple method which permits us to overcome the limits of previous analyses, thus providing a meaningful estimate of the UCER curve at the output of the concatenated system. The method has been applied to the relevant case of the new short LDPC codes recently proposed for updating the channel coding options in space TC links. We have shown that the codewords resulting from undetected error events at the output of the LDPC decoder very often have low weights. The CRC can reveal these low weight undetected error patterns, thus improving significantly the overall performance. This way, we have been able to confirm that, taking advantage of the CRC, the short LDPC codes can comply with the severe requirements set on the error detection capability for this kind of applications.
The main problem of the  proposed method is the difficulty in knowing the weight spectrum of the LDPC codes and the need to carry out very long simulations for estimating the UCER performance of the LDPC decoder.

\bibliographystyle{IEEEtran}
\bibliography{Archive}

\end{document}